\documentstyle[prb,aps,epsfig,multicol]{revtex}

\newcommand{\bg}[1]{{\mbox{$\mathversion{bold}#1$}}}

\newcommand{\cacuocl}{Ca$_2$CuO$_2$Cl$_2$ }
\newcommand{\mod}{\mathrm{mod}}
\newcommand{\SC}{\mathrm{SC}}
\newcommand{\DD}{two--dimensional }

\newcommand{\be}{\begin{equation}}
\newcommand{\ee}{\end{equation}}
\newcommand{\bea}{\begin{eqnarray}}
\newcommand{\eea}{\end{eqnarray}}
\newcommand{\n}{\nonumber \\}

\newcommand{\sgn}{\mathrm{sgn} }

\newcommand{\vk}{{\vec k}}

\newcommand{\eps}{\varepsilon}
\newcommand{\up}{\uparrow}
\newcommand{\down}{\downarrow}
\newcommand{\vkp}{{\vec k^\prime}}

\newcommand{\PSID}[1]{\Psi_{#1}^\dagger \,}
\newcommand{\gam}{\bg{\Gamma}}
\newcommand{\PSI}[1]{\Psi_{#1}^{} \,}
\newcommand{\sqmat}[4]
{\left( \begin{array}{cc} #1 & #2 \\ #3 & #4 \end{array} \right)}
\newcommand{\equ}[1]{equation (\ref{#1})}
\newcommand{\neel}{N\'{e}el }
\newcommand{\MF}{\mathrm{MF}}
\newcommand{\vrp}{{\vec r^\prime}}

\newcommand{\AF}{\mathrm{AF}}
\newcommand{\tvec}[2]
{\left( \begin{array}{c} #1 \\ #2 \end{array} \right)}

\newcommand{\SBMF}{\mathrm{SBMF}}

\renewcommand{\vr}{{\vec r}}

\def\v#1{{\bf #1}}

\renewcommand{\vec}[1]{{\bf #1}}

\def\al{\alpha}

\def\tc{T$_c$ }

\def\af{ antiferromagnetic }
\def\sc{ superconducting }

\def\sof{\hbox{SO$(5)$} }

\def\andword{and }
\def\submittedto{submitted to }
\def\toappearin{to appear in }

\def\percent{\%
}

\def\figref#1{{\protect\ref{#1}}}
\long\def\tagliasi#1{}

\newcommand{\cmag}{\gtrsim}

\def\up{\uparrow}
\def\down{\downarrow}

\def\beq{\begin{equation}}
\def\eeq{\end{equation}}
\def\beqn{\begin{eqnarray}}
\def\eeqn{\end{eqnarray}}

\def\eqref#1{ %  Eq.
 Eq. (\ref{#1})}

\def\andword{and }
\def\toappearin{to appear in }

\long\def\singlecol#1{
\twocolumn[\hsize\textwidth\columnwidth\hsize\csname @twocolumnfalse\endcsname
              #1]}

\ifpreprintsty %%c%%
\long\def\singlecol#1{#1}
\fi %%c%%

\def\bisco{Bi$_2$Sr$_2$CaCu$_2$O$_{8+\delta}$ }

\def\check#1{}

\global\firstfigfalse

\begin{document} 
\draft 
\title{
Interrelation between antiferromagnetic and superconducting gaps \\
in high-\tc materials 
%%%
}

\author{E. Arrigoni,  M. G. Zacher, T. Eckl, and W. Hanke}
%%%
\address{Institut f\"ur Theoretische Physik, Universit\"at W\"urzburg,
Am Hubland,  97074 W\"urzburg, Germany}
%%%
%%%

%%%
\maketitle

\begin{abstract}

We propose a 
phenomenological model, comprising a
microscopic \sof model plus the on-site Hubbard
interaction $U$
(``projected \sof model'') to understand
 the interrelation between the
$d$-wave-gap modulation observed by recent angle-resolved
photoemission experiments in the insulating antiferromagnet \cacuocl  
and the $d$-wave gap of high-\tc superconducting materials.
The on-site interaction $U$
is important in order to produce a
%%%
%%%
Mott gap of the correct
order of magnitude, which would be absent in an exact \sof theory.
 The projected \sof-model explains the gap
characteristics, namely both the symmetry and 
 the different order of magnitude of the gap
modulations between the \af and the \sc phases.
Furthermore, it is shown that the projected \sof theory can provide an explanation
for a recent observation [E. Pavarini et al., Phys. Rev. Lett. 87,
47003 (2001)], i. e. that the maximum \tc observed in a large variety
of high-\tc cuprates scales with the next-nearest-neighbor hopping
matrix element $t'$.
%%%
%%%
%%%
%%%
%%%

\end{abstract} 

\begin{multicols}{2}

\section{Introduction}
\label{intr}

The  \sof\ theory has been proposed as 
a unified description of antiferromagnetism (AF) and
$d$-wave superconductivity (SC) in the
high-\tc materials~\cite{zhan.97}.
While
originally formulated as a phenomenological effective field theory
for the global phase diagram of the high-\tc cuprates,
it was soon realized that the \sof symmetry,  in contrast to the 
more common spin SU(2) symmetry, is not realized exactly in 
 strongly-correlated models, 
such as the $t-J$ and Hubbard
model. However, it can be detected, under some circumstances,
 in the {\it low-energy sector} of such
 models~\cite{so5.97,ed.ha.97,me.ha.97,ha.ed.98.fp,ha.ed.98.pc,li.ba.98}.
This seems to indicate that this symmetry, which is broken at 
the microscopic level, may be restored, at least approximately, 
at long distances and low energies.

Recently, it was  pointed out that
an exact formulation of the \sof
theory cannot  properly account for the large Mott-insulating gap
which is present at half filling in the high-\tc cuprates:
an \sof transformation
 ``rotates'' spin into charge and, thus, 
an exactly \sof-invariant system 
should
have the same charge and spin gap, in contrast to the actual situation for
the high-\tc materials.
%%%
%%%
%%%
%%%
%%%
%%%
%%%
%%%
%%%
This prompted for an improved formulation of the
original idea  by introducing a
Gutzwiller 
``projection'' onto states where the constraint of no double occupancy 
is implemented locally~\cite{zh.hu.99,za.ha.00}. 
In particular, a  so-called ``projected'' \sof {\it bosonic} model
has been explicitly constructed~\cite{zh.hu.99}.
It is built up out of five bosons, three triplets which ``condense''
at low temperature into the \af state and two charged bosons,
i. e. Cooper pairs, which condense into the \sc state. The
``high-energy'' doubly occupied charged boson states are then
projected out from the Hilbert space. 
 The central hypothesis of the projected-\sof (p-\sof)
 theory is that the  model is accurate in describing both
the static and dynamic properties in the high-\tc superconductors (HTSC).

On the basis of a numerically exact Quantum-Monte-Carlo (QMC) calculation of
the p-\sof bosonic model, it has recently indeed been shown that this model gives a
realistic description of the phase diagram of the HTSC materials and
properly accounts for many of their physical properties\cite{do.ha.01u}.
Among those are the neutron-scattering resonance, which appears as a
Goldstone mode in the p-\sof description, as well as an unusual chemical
potential dependence on doping found in the prototype material 
$La_{2-x}Sr_xCuO_4$~\cite{in.mi.97},
which signals a possible (microscopic) phase separation\cite{we.le.97}. One
crucial point of this bosonic p-\sof model is that, on the basis of a newly
implemented Stochastic Series Expansion QMC technique, it can be simulated
up to unprecedented system sizes of about $10^4$ sites, which is more than one
order of magnitude larger than corresponding fermionic QMC simulations. This
fact allowed, for the first time, for an approximate ``finite-size-study'' of
the p-\sof bosonic model in the three-dimensional 
case, and for extracting the ``scaling
properties'' close to the bicritical (AF-SC) point. The numerical results
point to a partial asymptotic restoring of the \sof symmetry at this
critical point, i. e. at long distances. This is very
interesting, since the projection destroys the symmetry at the Hamiltonian
level (The Hamiltonian no longer commutes with the \sof generators). While
the mean-field classical p-\sof Hamiltonian of
Ref.~\onlinecite{zh.hu.99} 
conserves its \sof
invariant form, quantum fluctuations break the symmetry. However, as shown
in Ref.~\cite{ar.ha.00}, at finite temperature quantum fluctuations become less and
less important, and one can hope that symmetry-breaking effects due to the
projection become asymptotically irrelevant in the neighborhood of a
finite-temperature critical point, which is what the numerical results seem
to find.

Therefore, one part of our motivation to study the p-\sof model on the
{\it fermionic level} stems from the success of the bosonic model. Thus, in the
present paper, we start again from the assumption that the high-\tc materials
are characterized by an underlying approximate \sof symmetry. This
assumption  is further
supplemented by the Gutzwiller ``projection'', i. e. the
Hubbard-$U$-term provides
 an
important additional ingredient to the \sof-symmetric part of the
Hamiltonian, which guarantees the presence of a Mott gap at
half-filling. 
In
this sense, we choose our Hamiltonian as the simplest (namely scalar \sof
invariant~\cite{ra.ko.97})
 Hamiltonian, which possesses such a symmetry and which
displays the correct d-wave superconducting gap at finite doping. The
physical basis is thus the assumption of a {\it projected} \sof symmetry, and the
existence of a superconducting state at finite doping. The $t-J$ model was
shown by our previous numerical studies~\cite{ed.ha.97}
 to have (nearly) projected \sof
symmetry in the bosonic sector. However, the $t-J$ model cannot explain the
$|d|$-wave AF gap modulation in the fermionic sector, it therefore misses an
important piece of physics our current model contains (unless one introduces
{\it ad-hoc}
 values for longer-ranged hoppings $t', t''$, see below). Thus, the
logic of our approach is to construct an effective model, which matches the
low-energy experiments, rather than trying to derive an effective model from
first principle. This is the central idea behind a Landau-type approach to
strongly-correlated systems, and can be very useful, although not complete.

This brings us to the second part of our motivation. It is the relation to
recent angle-resolved photoemission data, which found evidence for a
$d$-wave-like modulation of the antiferromagnetic gap, suggesting an intimate
interrelation between the AF insulator and the SC with its $d$-wave gap. In a
recent letter~\cite{za.ha.00} it was shown, that the projected \sof theory correctly
describes the observed gap characteristics. Specifically, it accounts for
the order of magnitude difference between the AF gap modulation of order
$J\approx 0.2$ eV) and the SC gap ($\sim J/10$) and is also consistent with the $d$-wave gap
dispersion. Thus the projection is crucial, since in an exactly
\sof-invariant model, the $d$-wave SC gap would be transformed into a pure
$|d|$-wave AF gap with the same amplitude and without a constant ($s$-wave)
part. In 
Ref.~\onlinecite{za.ha.00}, the on-site Hubbard $U$ has been treated by a Hartree-Fock
decoupling, unifying the gap interrelation in the much-used
spin-density-wave (SDW) picture for the AF and BCS limit for the SC. This
treatment will be relegated here to a short appendix (Sec.~\ref{appa}), introducing some
notations and also giving a simple illustration of the crucial difference
the ``projection'', i. e. the $U$-term induces in the AF gap modulation and in
its SC counterpart. In the present paper, we concentrate on a more
controlled calculation for larger $U$ ($U\gg t$), i.e. a slave-boson calculation.

This paper is organized as follows:

In Sec.~\ref{expe},
we start with a summary of the 
experimental results observed by angular-resolved photoemission
spectroscopy (ARPES), which have motivated our
calculation.  
In Sec.~\ref{mode}, we show that the $\pi$-operators of the \sof
symmetry ``rotate'' the $d$-wave symmetry of the SC order parameter into the
$|d|$-wave symmetry of the AF order parameter. This argument is general, and
independent of the details (in particular, it survives the projection). More
specifically then, we discuss the possibility of introducing a generalized
form factor $g_k$ for the \sof symmetry, which emphasizes the importance of
longer-ranged pairings in accordance with the experimental observations.
The comparison of our results with ARPES experiments is given in
Sec.~\ref{disc}.
In Sec.~\ref{slav} we compare the mean-field result with a
 slave-boson approach.
The treatment of the 
antiferromagnetic and of the superconducting phase are given in
Sec.~\ref{sbaf} and \ref{sbsc}, respectively.
%%%
The usually employed $t'$-($t''$) fitting procedure in
Hubbard-($t-J$-) like models to ARPES data seems, at first sight, to be
completely uncorrelated with the experimental finding, which emphasizes the
universal role of the magnetic energy scale $J$, namely the fact that the
insulating bandwidth itself is of order $J$. In Sec. \ref{effe}, we
show that the $t'-t''$ -hoppings can indeed arise effectively from the \sof part of the
Hamiltonian, i. e. from the spin interaction term. Since in the \sof
description these hoppings scale with the SC strength ($V$ in Eq.~\ref{hsc} ), they
are also qualitatively in line, from a rather different point of view, with
a recent observation~\cite{pa.da.01} i.e. that the maximum \tc observed in a
variety of HTSC scale with the next-nearest-neighbor hopping matrix element $t'$.
This empirical observation and its p-\sof interpretation will be discussed
in Sec. \ref{effe}.
%%%
%%%
Finally,  Sec.~\ref{summ} gives a summary of our results.

%%%
\section{Experimental Guidelines}
%%%
\label{expe}

Our analysis is motivated by
 recent ARPES
%%%
experiments by the Stanford 
group\cite{ro.ki.98},
which indicate
an intimate interrelation between the
superconducting and antiferromagnetic state.
These experiments
show remnants of a superconducting gap in the 
half--filled high-\tc parent compound \cacuocl. 
The ARPES gap structure in the AF phase  is
given by the dots with error bars in Fig.~1 of
Ref.~\onlinecite{za.ha.00}. It
 displays a $d$--wave--like, i.e. 
$|\cos k_x - \cos k_y|$--like dispersion in the one--electron spectral
function $A(\vec k, 
\omega)$ with respect to the lowest energy state at wave--vector $\vec
k = (\pi/2,\pi/2)$.  
The points
 show a $d$--wave dispersion function 
of the photoemission band
along the
edge of the magnetic Brillouin zone, which, by assuming a symmetric 
inverse photoemission band, can be understood as a modulation
of the single-particle Mott-Hubbard gap.
The inset of this figure presents the ARPES data and the dispersion
 features in a \DD
plot: on a line drawn from the center of the Brillouin zone to the experimental points, the
distance between these points to the intersection of this line with the AF Brillouin
zone gives the value of the dispersion at the $\vec k$--point
considered. The data closely follow the 
$|\cos k_{x}-\cos k_{y}|$--behavior, 
depicted in the $d$--wave full line.
 The important point is that
these photoemission data suggest that the $d$--wave--like dispersion in the
insulator is also the underlying reason for the  pseudogap in the
underdoped regime: this ''high--energy'' pseudogap of the order $J\sim 0.1eV$
continuously evolves out of the insulating feature, as documented not only
by the same energy scale but again by the same $d$--wave dispersion
\cite{ma.de.96}. 
Since, on the other hand, 
this high--energy feature is  closely correlated to the superconducting gap
as a function of both
doping and momentum\cite{wh.sh.96,ma.de.96,di.yo.96},
we conclude that a microscopic theory
should be able to explain the
interrelation between the superconducting gap and the AF gap modulation.
In Ref.~\onlinecite{za.ha.00}, we have shown that a  \sof model plus $U$ 
provides such an interrelation in a natural way.
In the present paper, we want to discuss this issue in more detail.

%%%
\section{Microscopic model}
%%%
\label{mode}

\begin{figure}
\begin{center}
\epsfig{file=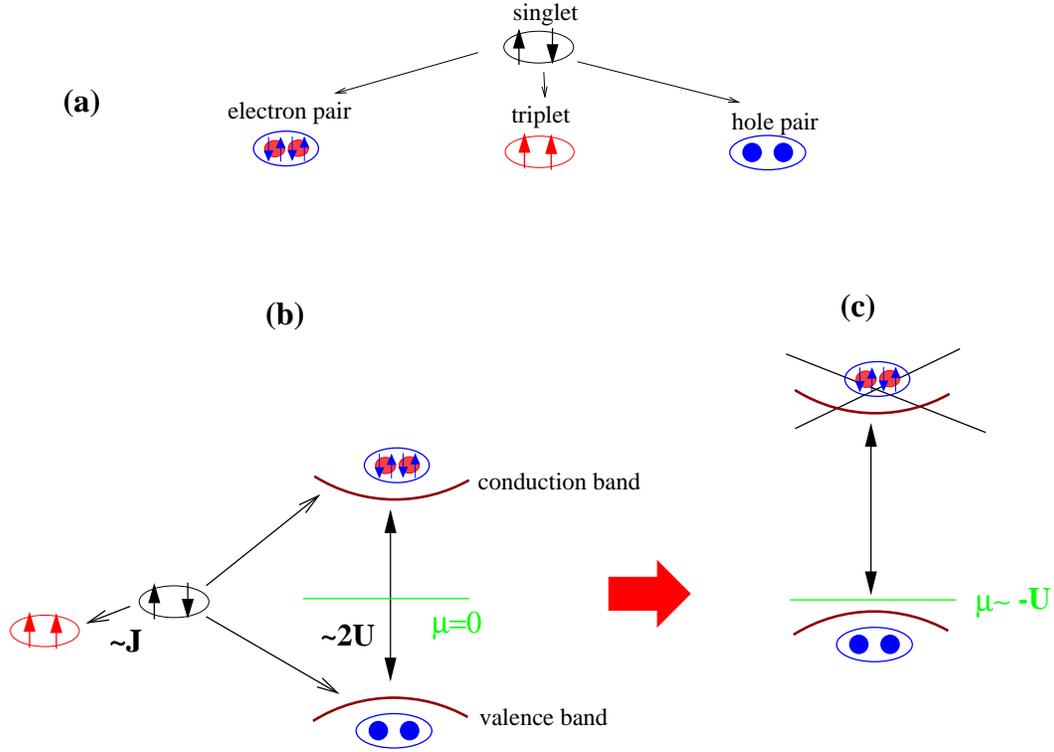, width=14cm}
\vspace*{1cm}
\caption{
Suppose one
starts from 
some  singlet
state~\cite{zhan.97,eder.99} 
 as a description of the vacuum [e.g. a resonating-valence-bond-like (RVB)
 state],
 and looks
for excitations on top of it. 
In an (ideal) exact \sof--symmetric description (a), 
triplet--, hole--pair-- and electron--pair--excitations
on top of this RVB state
are equivalent.
 This condition is obviously violated in 
the cuprate materials at half filling
where spin--excitations are gapless (due to the \af state), while
 hole--/electron--pair 
excitations have a large gap of the order 
of the Mott gap (b).
A projected \sof theory, however, can restore 
the 
equivalence between  spin and charge excitations 
{\it in the hole sector}. This is done 
by 
lowering the chemical potential to the top of the valence band, and
projecting out high-energy electron-pair excitations
so that 
 hole--pair excitations become gapless (c). 
}
\label{figenricoproj}
\end{center}
\end{figure}

As discussed above,
an  important point missing in an exact \sof theory
is the correct description of the Mott--gap at half--filling.
For this reason, it is important 
to account for the strong on--site Coulomb repulsion $U$.
%%%
%%%
This can  naturally be
 done by including the usual Hubbard on-site interaction term $U$ ``by hand''.
Formally, one can project out  
 doubly-occupied sites by taking $U\to \infty$, eventually.
This should be a
 good approximation since these states are separated by an energy
scale,  which is orders of magnitude higher than the interesting
low--energy scales set by the N\'eel ($T_N$) and by the
superconducting ($T_c$) transition temperatures.
The idea of 
 such a ''projected'' \sof theory, 
recently 
formulated in a rigorous fashion for {\it bosonic} degrees of 
freedom in Ref.~\onlinecite{zh.hu.99}, 
 is
visualized in Fig.~\ref{figenricoproj}.

As already mentioned in the Introduction, the physical basis for the
choice of our Hamiltonian is the {\it assumption} that the high-\tc
materials are following an underlying \sof symmetry.
The hamiltonian in Eqs.~(\ref{hso5}-\ref{htot}) has been chosen as the
simplest \sof-symmetric hamiltonian that can properly reproduce the
$d$-wave \sc state of the High-\tc materials in a BCS mean-field
description. Our starting point is the \sc state. We then perform an
\sof rotation on the operator level that introduces the magnetic part
of the interaction [second term in Eq.~\ref{hso5}]. The resulting
hamiltonian is then \sof invariant. This hamiltonian must be further
supplemented with the projection (Hubbard-$U$) term and the usual
chemical potential term.

Our aim is to describe 
 the peculiar ARPES results of
Ref.~\onlinecite{ro.ki.98}
 by the simple assumption of \sof symmetry of the
microscopic Hamiltonian {\it plus} the projection prescription. 
The first task is, thus, to build up the {\it simplest} fermionic
lattice Hamiltonian obeying these two requirements.
%%%
%%%
%%%
The part of the Hamiltonian leading to the superconducting state 
is dictated by experiments on the SC gap in the cuprates. These
experiments suggest  
a $d$--wave superconducting gap resulting in 
 a nearest neighbor $d$--wave BCS interaction.
We also allow for more general forms of the SC gap function,
including longer-range interactions, as observed in more accurate
ARPES experiments on \bisco~\cite{me.no.99}.
With the SC part of the Hamiltonian being fixed, the AF part 
is then also fixed
 by the \sof rotation.
Therefore, once 
the form of the wavefunction in the SC state is given, {\it the \sof
assumption allows to 
 make a prediction about the AF state}.

\end{multicols}
%%%
\begin{multicols}{2}

To see this, let us consider the operator creating a  Cooper pair with 
vanishing total momentum: 
%%%
\beq
\label{P}
  P(\vec k) \equiv \frac{i}{2} \ c_{\vec k}\,  \sigma^y \, c_{-\vec k} \;,
\eeq
where $c_{\vec k} = ( c_{\vec k,\up} , c_{\vec k,\down} )  $ is a
two-component destruction operator for 
an electron with 
momentum $\vec k$, and
$\sigma^{\al}$ are the usual Pauli matrices ($\al=x,y,z$).
An \sof rotation is generated by the $\pi$
operator~\cite{zhan.97,ra.ko.97,henl.97} 
\beq
\label{pial}
\pi_{\al} = \sum_{\v k} g_{\vec k} \ c_{\vec k + \vec Q} \ 
\sigma^{\al} \sigma^y \ c_{- \vec k} \;,
\eeq
with the form factor
\beq
\label{gnn}
  g_{\vec k} = \sgn (\cos k_x - \cos k_y) \;.
\eeq
The $\pi$ operator
 rotates the Cooper pair  into
the magnon operator
\beq
\label{N}
  N_\alpha ({\vec k}) \equiv \frac{1}{2}\ 
 c^\dagger_{\vec k + \vec Q} \, \sigma^\alpha \, c^{}_{\vec k} \;,
\eeq
via the commutation rules
\beq
\label{commn}
\Big[ \pi_\alpha, N_\beta(\vec k) \Big]  =  -i \ \delta_{\alpha,\beta}\,
g_{\vec k}\,  P(\vec k) \;,
\eeq
and
\beq
\label{commp}
 \Big[ \pi_\alpha^\dagger, P(\vec k) \Big] = i \ g_{\vec k} \,
 N_\alpha(\vec k) \;.
\eeq
From \eqref{commn} and \eqref{commp} it is clear that a $d$-wave
Cooper pair $d_{\vec k} P(\vec k)$
with the nearest-neighbor
gap
dispersion
\beq
\label{dnn}
d_{\vec k} = (\cos k_x - \cos k_y) \;.
\eeq
transforms into a ``$|d|$-wave-shape''
magnon $|d_{\vec k}| N_{\beta}(\vec k)$, giving a corresponding
``$|d|$-wave'' gap in the AF state.
In its original formulation, an \sof rotation was defined by the $\pi$ 
operator, \eqref{pial}, with the ``nearest-neighbor'' form factor \eqref{gnn}.
From the above discussion it is clear that such a formulation is only
appropriate for a similarly nearest-neighbor \sc
gap function, \eqref{dnn}. An extended gap, as found in
recent ARPES experiments on 
\bisco~\cite{me.no.99} would not be transformed into 
the appropriate AF form by this \sof transformation.
However, \eqref{gnn} is not the only possible form of the form
factor which can be used.
In fact,
 one can use a {\it different} \sof symmetry
transformation with a {\it more general} form of $g_{\vec  k}$.
More specifically, it can be shown that the $\pi$ operator
\eqref{pial} satisfies
the closure requirement of the 
\sof group 
 as long as 
the form factor has the properties
$|g_{\vec  k}|=1$, and $g_{\vec  k+\vec Q}=-g_{- \vec k}$.
Thus, given a more general form of the gap dispersion
$d_{\vec k}$ satisfying $d_{\vec k+\vec Q}= - d_{-\vec k}$, one can choose
\beq
\label{ggen}
g_{\vec k} = \sgn \ d_{\vec k} \;.
\eeq
A gap function satisfying these requirement can support, e. g.,
third-neighbor terms:
\beq
\label{dgen}
d_{\vec k} =  b (\cos k_x - \cos k_y) + (1-b) 
(\cos 3 k_x - \cos 3 k_y) 
\;.
\eeq
From the commutators, \eqref{commn} and \eqref{commp}, it is thus clear
that  the generalized choice \eqref{ggen}
 produces a corresponding AF gap function 
of the form
$|d_{\vec k}|$.

As explained above, our procedure starts by taking the simplest
fermionic lattice Hamiltonian reproducing the appropriate $d$-wave
state in the SC state within a BCS (mean-field) description.
 This is given by 
\beq
\label{hsc}
H_{SC} = 
H_{kin} +
\frac{V}{2 N} \left[
\Delta \Delta^{\dag} + \Delta^{\dag} \Delta
\right] \;,
\eeq
%%%
%%%
%%%
where the kinetic-energy term reads
\beq
H_{kin} =
\sum_{p,\sigma }\varepsilon_{p} c_{p,\sigma}^{\dagger} c_{p,\sigma } \;.
\eeq
Here, $\varepsilon_p$ is
 the  single-particle kinetic energy, and
$\Delta$ 
the $d$-wave pairing operator 
\beq
\label{Delta}
\Delta \equiv \sum_{\vec k} d_{\vec k} P(\vec k) \;.
%%%
\eeq
with $P(\vec k)$ given by \eqref{P}.

The second term on the right-hand side 
of \eqref{hsc} 
describes a coupling between
Cooper pairs. The \sof assumption requires 
that, (i), a corresponding 
coupling between magnons be present as well, which leads to the form
\beq
\label{hso5}
H_{\sof} = H_{SC} - 
\frac{V}{N} \vec m \cdot \vec m \;,
\eeq
where
\beq
\label{mal}
 m_{\al} \equiv \sum_{\vec k} |d_{\vec k}| N_{\al}(\vec k) \;,
\eeq
and, (ii), that $\varepsilon_{p+Q} = - \varepsilon_{-p}$, which suggests 
the simple nearest-neighbor form 
$\varepsilon_p = - 2 t (\cos p_x + \cos p_y)$.

\begin{figure}
\begin{center}
\epsfig{file=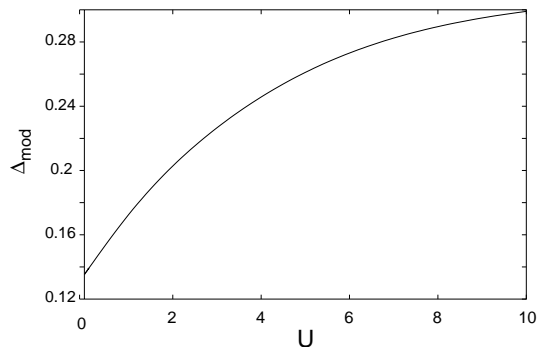, height=7cm, angle=270}
\vspace*{4mm}
%%%
\caption{The amplitude $\Delta_{\textrm{mod}}$ of the AF 
$d$--wave--like modulation 
as a 
function of the Hubbard--interaction $U$ for fixed 
$V=0.61 \ t$ and $b=1$.
The figure plots the results of 
 the slave-boson calculation of Sec.~\figref{slav}.
}
\label{fgapafu}
\end{center}
\end{figure}

As discussed in the Introduction, restricting to a purely \sof-symmetric 
Hamiltonian like \eqref{hso5}, would produce a gap with nodes in the AF
state and no constant part of the Mott-insulating gap, as present in
the high-\tc materials.
For this reason, our Hamiltonian must be supplemented  
with a term suppressing double occupation at half filling.
This can be achieved by
introducing 
the usual on-site Hubbard-repulsion term
\beq
\label{hu}
  H_{U} =
U \sum_{\vec r} \left( n_{\vec r,\up} n_{\vec r,\down} 
- \frac12 \sum_{\sigma} n_{\vec r,\sigma} \right)
\, ,
\eeq
where, 
$ 
n_{\vec r,\sigma} \equiv
 c^\dagger_{\vec r, \sigma} c^{}_{\vec r, \sigma}
$ 
is the  particle number with spin $\sigma$ on site $\vec r$.
and
for convenience, 
we have subtracted a constant chemical-potential term.
Clearly, for a {\it complete} suppression one should take 
 the $U \to \infty$ limit. However, here we will
consider a finite $U=8 \ t$ giving 
a more physical {\it partial} projection
resulting in a finite Mott gap at
half filling.
It is clear that,  concerning the
 energy scales of the order
$J$ we are interested in, there is very little difference between
results at $U/t=8 
$, and $U/t=\infty$. This will be confirmed by our calculation later,
see Fig.~\ref{fgapafu}.
%%%
%%%
The last term in the Hamiltonian is the
usual chemical potential term
$H_\mu$  controlling the doping, i. e.
\beq
\label{hmu}
  H_\mu = - \mu \sum_{\vec r,\sigma} c^\dagger_{\vec r, \sigma} 
c^{}_{\vec r, \sigma} \;.
\eeq
In total, our Hamiltonian has the form
\beq
\label{htot}
H_{tot} = H_{\sof} + H_U + H_{\mu} \;.
\eeq
In Ref.~\onlinecite{ar.ha.00}, we have shown that the chemical
potential term partially compensates for the \sof symmetry breaking of 
the Hubbard term $H_U$ (see also schematic Fig.~\ref{figenricoproj}c), 
as also shown for the $t-J$ model in Ref.~\onlinecite{ed.ha.97}.

The interacting part of $H_{\sof}$ is a special case of the manifestly 
\sof invariant part introduced by Rabello et al.~\cite{ra.ko.97}:
\beqn
\label{equsofint}
&&  H_{\sof,I} \equiv H_{\sof}-H_{kin} =  \sum_{a=1}^5
   \sum_{\vk,\vkp,\vec q} V_1(\vk, \vkp, \vec q)
   \, 
\nonumber \\ && \times
   \left( \PSID{\vk} \gam^a \, \PSI{\vk+\vec q} \right) 
   \left( \PSID{\vkp} \gam^a \, \PSI{\vkp-\vec q} \right) \, ,
\eeqn
where 
\beq 
  \PSI{\vk} = \left\{ c_{\vk,\uparrow}^{}, \, c_{\vk,\downarrow}^{},  \,
                 g(\vec k) c_{-\vk+\vec Q,\uparrow}^\dagger, \,
                 g(\vec k) c_{-\vk+\vec Q,\downarrow}^\dagger \right\} \, 
,
\eeq
is an \sof spinor, and the $\gam$ matrices 
\bea
  \bg{\Gamma}^1 & = & \sqmat{0}{-i \bg{\sigma}^y}{i \bg{\sigma}^y}{0},  \quad
    \bg{\Gamma}^{2,3,4} = \sqmat{\sigma^{x/y/z}}{0}{0}{{\sigma^{x/y/z}}^T} \n
  \bg{\Gamma}^5 & = & \sqmat{0}{\bg{\sigma^y}}{\bg{\sigma^y}}{0} \;,
\eea
provide the \sof group structure.
In particular, our exactly \sof invariant part of the Hamiltonian, 
\eqref{hso5} is
given by \eqref{equsofint}
with the simple choice
\beq
V_1(\vk, \vkp, \vec q) = 
     - \frac{V}{16 \ N} 
 \, |d_{\vk}| \, |d_{\vkp}| \, \delta_{\vec q,\vec Q} \, , 
\eeq

Our ``minimal'' projected \sof model 
contains, in principle, five parameters, although only two of them are
adjustable.  
The adjustable parameters are the interaction term $V$, and $b$ 
[cf. \eqref{dgen}].
 $V$, due to the \sof condition,
 controls the magnitude of both the interactions between
Cooper pairs and between magnons.
As anticipated, we fix this parameter by fitting the experimental
magnitude of
the gap in the SC phase.
The second one, $b$ , is needed when the SC gap deviates 
from the nearest-neighbor
 form. This parameter can be fitted by the detailed form
of the  gap  in the SC phase.
The other  three parameters, $\mu$, $t$, and $U/t$, are not free.
$\mu$ is fixed by the hole density, which we take from the experiments 
to be $<n>  = 0.89$~\cite{me.no.99} in the SC phase,
while for $t$ we take the commonly accepted value
 $t= 0.5 eV$.
Several experiments and theoretical fittings agree
about a strong-coupling value of $U$ of the order $U/t\approx 8$.
On the other hand, we show that our results saturate in the large-$U$
limit, i. e. they are practically independent of $U$ for $U/t \cmag 8$.

We now proceed to illustrate the mean-field treatment of our
Hamiltonian \eqref{htot}.
The main point of this treatment is to (a) show that the
$|d|$-modulation survives the Gutzwiller projection. This important
property further clarifies what is meant by ``projected \sof
symmetry'' in the fermionic sector, generalizing the concept
introduced in Ref.~\cite{zh.hu.99}. Furthermore, we show (b) that the
\sof projection introduces a difference in the magnitude of the \sc
and \af gap modulation which is consistent with the experiment. This
quantitative analysis is more model dependent compared to point (a),
but within the approximations we show the robustness of the result.

Details of the
SDW treatment of the \af phase and of the 
 BCS treatment of the \sc phase
as well as some basic notations can be found in the appendix, Sec.~\ref{appa}.

\subsection{Discussion of the SDW solution}

\begin{figure}
\begin{center}
\epsfig{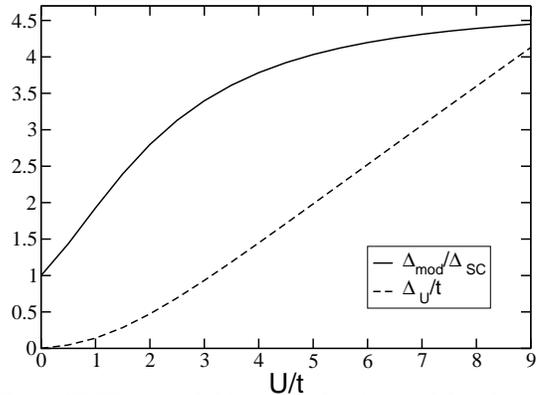}
%%%
\caption{SDW mean-field results for 
%%%
%%%
$\Delta_U$ 
 and for the ratio of the 
$d$--wave--like modulated parts of the gap $\Delta_{\mod}/\Delta_{\SC}$
(both at half filling) as a function of $U$. }
\label{figgapaf}
\end{center}
\end{figure}

\eqref{deltaudeltamod}
yields the coupled set of self--consistent equations:
\bea
\label{eqaf}
\Delta_{\mod} &=& \frac{ V}{2 \, N} \sum_{\vk} |d_{\vk}| 
                \frac{\Delta_{\mod} |d_{\vk}| + \Delta_U}{E(\vk)} \, , \\
\Delta_U &=& \frac{U}{2 \, N} \sum_{\vk} 
                \frac{\Delta_{\mod} |d_{\vk}| + \Delta_U}{E(\vk)} \, ,
\eea
where $E(\vk)$ is the quasiparticle energy given in Eq.~\ref{ek}, and
where the sum runs again over the {\it whole} Brillouin zone.
The behavior of $\Delta_U$ and of ratio $\Delta_{mod}/\Delta_{SC}$ 
are reported in Fig.~\ref{figgapaf}.
As one can see, the ratio between the AF and SC gaps, which is unity at
the exactly \sof-symmetric
 point $U/t=0$, increases with $U$ and saturates at
$U/t\approx 8$.

%%%
%%%

\begin{figure} \begin{center}
\epsfig{file=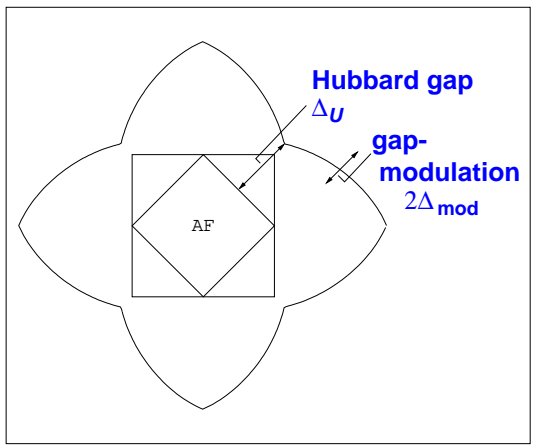, width=7cm}
\vspace*{.5cm}
\caption{
Schematic gap structure in the AF phase along the magnetic zone boundary.
On a line drawn from the center of the Brillouin zone to any point on
the curve, the distance from this point to the intersection of the
line with the antiferromagnetic 
Brillouin zone boundary gives 
the gap at the $\vec k$--point of the intersection.
On top of the constant Hubbard gap, 
which, for convenience, has been
 reduced in scale, there is a $d$-wave like modulation. 
}
\label{figafgap}
\end{center} \end{figure}

We now study the gap structure of the single--particle spectrum along the magnetic zone
boundary. Since 
 the free dispersion vanishes here, the gap becomes
\bea
\label{deltaaf}
  \Delta_{\AF}(\vk) &=& E^v(\vk) - E^c(\vk) \n
  &=& 2 (\Delta_U + |d_{\vk}| \, \Delta_{\mod}) 
 \, .
\eea
 We, thus, obtain
a
$d$--wave--like modulation with amplitude
$2 \, \Delta_{\mod}$ on top of the constant Hubbard gap $2 \,
\Delta_U$, as anticipated above.
This gap structure is represented schematically in Fig.~(\ref{figafgap}).
%%%
%%%

%%%
\subsection{Physical interpretation of the \sof-generated
  (long-range) magnetic interaction part}
%%%
\label{effe}

It is instructive to transform 
the magnon term \eqref{mal} entering 
the magnetic part of the interaction (last term on the r.h.s. of
 \eqref{hso5}) into real space. This gives
\beq
\vec m = \sum_{\vec r} \vec m(\vec r) \;,
\eeq
where the N\'eel order parameter at site $\vec r$ reads 
\beq
m_{\al}(\vec r) = \frac12 \sum_{\vec r_2}  w(\vec r-\vec r_2)  \, 
              e^{i \vec Q \vec r_2}  \,\, c^\dagger_{\vec r} \, 
              \sigma^{\al}  \, c_{\vec r_2} \, ,
\eeq
and
has an extended internal structure 
given by the
 Fourier transform 
$w(\vec r)$
of 
$|d_{\vec k}|$.
This function is given by
(cf. Ref.\onlinecite{ra.ko.97})
\beq
 w(m \, \vec a_x + n \, \vec a_y) = \frac{4}{\pi^2} \frac{1 + (-1)^{m+n}}
    {[(m+n)^2 - 1] [ (m-n)^2 -1]} \, .
\eeq
We want to show 
 how, at the mean-field level,  
the extended structure of the magnetic part of the \sof--symmetric
Hamiltonian leads to
effective longer--ranged hopping terms 
similar to the  commonly used values.
These terms
can be explicitly seen in the
mean-field \sof--symmetric Hamiltonian in real space

\be
[H_{\sof}]_{{MF}} = 
- \Delta_{\mod} \sum_{\vr, \vrp,\sigma} \sigma w(\vr - \vrp) e^{i \vec Q \vr}
  c^\dagger_{\vr,\sigma} c^{}_{\vrp,\sigma} + H_{SC} \;.
  \label{equmagmean}
\ee
Notice that  
the effective hopping terms only connect sites within the same sublattice,
since $ w(\vr - \vrp)$ vanishes otherwise.
 Furthermore, the sign of the effective hopping matrix
elements is opposite for spin up and down, as well as
 for the two sublattices, 
thus yielding an
identical sign for the majority spin species on each sublattice
i.e. for the electrons whose spin is polarized
along the direction of the staggered magnetic field. The processes
 involving those
''majority'' electrons are the most important ones since 
the ratio between the number of ''major'' to ''minor'' electrons is
\be
  N_{\mathrm{major}}/N_{\mathrm{minor}} = 
  (1 + \langle m\rangle)/(1 - \langle m \rangle) \, ,
\ee
where $\langle m \rangle = \frac{\Delta_U}{U}$ is the mean-field 
staggered magnetization which is of
the order $\langle m \rangle \approx \frac{1}{2}$.

The effective second--
and third--nearest neighbor ''hopping amplitudes''
 generated by the interaction \eqref{equmagmean} are
\bea
\label{ttp}
  t^\prime &=& \Delta_{\mod} \frac{-8}{3 \pi^2} \approx -0.08 t\, , \\
  t^{\prime\prime} &=& \Delta_{\mod} \frac{8}{9 \pi^2} \approx +0.03 t\, , \\ 
  t^{\prime\prime}/t^\prime &=& - 1/3 \, .
\eea
Here we have used the self--consistent solution  $\Delta_{\mod}
\approx 0.3 t$  
for large $U$ 
and the results of Sec.~\ref{hart}.
These effective
second-- and third--nearest neighbor hopping elements
have the same sign but 
are somewhat smaller than 
 the ones
commonly introduced in order to correctly reproduce
the Fermi-surface topology of typical
high-\tc materials (see, e. g., Ref.~\onlinecite{ki.wh.98}).
%%%
Nevertheless,
here, these ``effective'' parameters are obtained 
without any fitting to the experiments by the simple 
 assumption that leads us to the projected \sof Hamiltonian \eqref{htot}.
This makes clear, in particular, that the gap modulation is related to 
the magnetic energy scale $J$, as demonstrated in the ARPES
experiments and repeatedly stressed in the literature~\cite{laug.97}.
\eqref{equmagmean}
also contains  
longer--ranged hopping
processes which, however, are much smaller than $t'$ and $t''$.
 These long-range terms, however, produce 
 the cusp--like feature 
of the antiferromagnetic gap modulation which has been observed in
ARPES experiments 
 on \cacuocl \cite{ro.ki.98}.

\begin{figure} \begin{center}
\epsfig{file=tcts-new.eps, width=7cm} \\[.4cm]
\epsfig{file=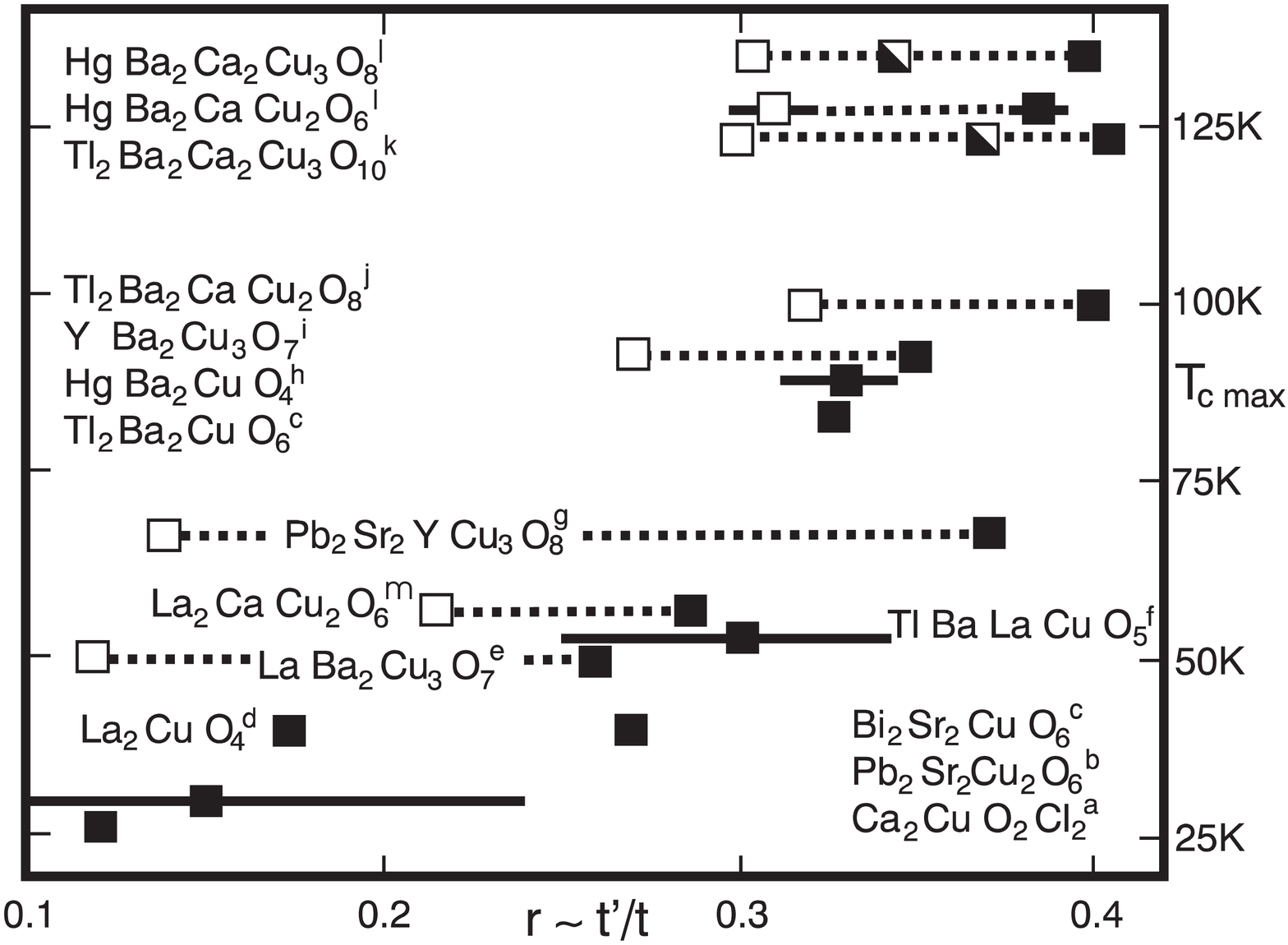,angle=-90,width=7cm}
\vspace*{.5cm}
\caption{
\label{tcts}
Correlation between $T_{c}$ and the next-nearest-neighbor hopping
$t'$ (in units of $t$).
The top figure reports the results of our calculation for, whereby
different $t'$ are obtained by varying $V$, and the superconducting
\tc is evaluated via the weak-coupling BCS equation. 
The doping is fixed at $15\percent$ ($<n>=0.85$),  $U=8t$, and $b=1$.
For comparison,
 the bottom figure  shows the results
of Ref.~\protect\onlinecite{pa.da.01} for diffent materials.
}
\end{center} \end{figure}

Since  the hoppings \eqref{ttp}
scale with the strength of the \sof coupling
($V$ in Eq.~\ref{hsc}),
and since, at least in the weak-coupling regime $V\ll 8t$ 
the \sc \tc increases with $V$, one can expect that
\tc is an increasing function of the next-nearest-neighbor hopping $t'$. 
This fact has been observed in a recent analysis of Pavarini {\it et
  al.}\cite{pa.da.01}, who have plotted the maximum \tc as a function of the ratio
$t'/t$ for different compounds.
As explained above, our theory provides a {\it qualitative}
explanation for such a behavior. This is more quantitatively 
 shown in Fig.~\ref{tcts}, which plots our \tc versus $t'$ results in
 the upper part.
Here, different next-neares-neighbor hoppings $t'$ have been extracted
by varying the \sof coupling $V$. For a given $V$, and thus
$t'$, \tc has then been extracted from the weak-coupling BCS equation,
which is valid for $V\lesssim 8t$.
The lower part of Fig.~\ref{tcts} gives finally the \tc versus $t'$
results of Pavarini {\it et al.}\cite{pa.da.01} for a large variety of
HTSC materials.

\subsection{Discussion of ARPES experiments}
\label{disc}
The interesting point is that the
structure of the AF gap \eqref{deltaaf}, shown in Fig.~\ref{figafgap}
 is a consequence of the \sof-symmetry principle combined with the
effect of the Hubbard interaction $U$.
 This 
gap
structure  has recently 
been  measured in
the undoped (half--filled) cuprate parent compound \cacuocl \cite{ro.ki.98}.
Fig.~1 of Ref.~\onlinecite{za.ha.00} shows the ARPES results 
for the AF gap modulation
on this material, in comparison with our result, \eqref{deltaaf}.
%%%
%%%
%%%
%%%
%%%
%%%
%%%
%%%
%%%
%%%
 The  experimental
 figure displays
 the peak dispersion 
versus the
$d$--wave--like function $|\cos k_x - \cos k_y|$.
The very good 
fit to a straight line
shows that
the dispersion is $d$--wave--like with a characteristic ''cusp'' at momentum 
$\vec k =(\pi/2,\pi/2)$.

\begin{figure}
\begin{center}
\epsfig{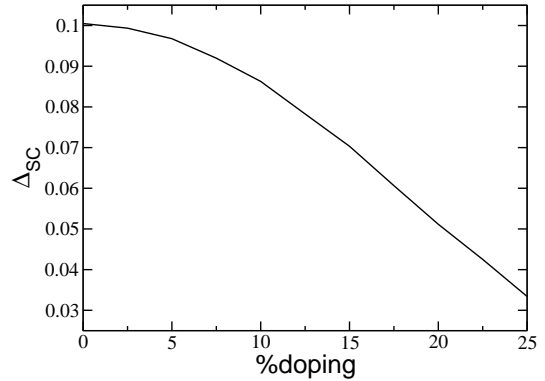}
\vspace*{4mm}
%%%
\caption{Mean-field result for 
the $d$-wave SC gap $\Delta_{\SC}$ obtained by solving the gap equation 
(\ref{equbcsgap}) as a function of
doping 
($1-\langle n \rangle$) for $V = 0.8 t$ and $b=1$. 
At a doping of $\approx 20 \%$,
$\Delta_{\SC}$ takes the value \mbox{$\Delta_{\SC}\approx 0.05 t$}
which is about the correct order of
magnitude for the high-\tc materials ($\Delta_{\SC} \approx J/10$).}
\label{figgapscofn}
\end{center}
 \end{figure}

Concerning the energy scales that are observed in the
experiment~\cite{ro.ki.98},  the
authors find a 
modulation $\Delta_{\mod}$ of the order the magnetic exchange coupling $J$
which is about one order of magnitude higher than the 
experimentally observed amplitude of the superconducting gap
$\Delta_{\SC} \approx J/10$. 
This is a second, most important point that would be difficult to
understand in terms of an ``exact'' \sof symmetry, which would 
be expected to preserve
the amplitude of the modulation by ``rotating'' from the SC to
the AF gap.
However, the  result of our {\it projected} \sof calculation 
is that the 
difference 
by one order of magnitude between the two amplitudes
is correctly 
reproduced, and it is due to the projection as well.
This can be seen from
Fig.~(\ref{figgapaf}), displaying
 the ratio $\Delta_{\mod}/\Delta_{\SC}$
as obtained by our mean-field calculation 
as a function of the Hubbard repulsion $U$. We see that 
for large $U \approx 8 t$
the AF gap modulation $\Delta_{\mod}$ is 
approximately four times larger 
 than the
superconducting gap $\Delta_{\SC}$, when the latter is also calculated at
half filling. 
An additional factor
two emerges from the doping dependence of $\Delta_{\SC}$ shown 
in Fig.~(\ref{figgapscofn}) so that the 
total ratio of $\Delta_{\mod}/\Delta_{\SC}$ becomes of the order of
$8$.

%%%
%%%

At first sight, it might seem difficult to 
 understand why the introduction of $U$ changes the order of
magnitude of the SC and AF gap modulations,
despite the fact that
 both are controlled by the
same energy scale $V$. This can be easily understand at weak
coupling $V\ll t$ and for $U\gg t$. 
In this case, the usual  BCS gap equation  \eqref{equbcsgap} -- which
does not depend on $U$ -- gives
$\Delta_{SC} \sim \exp -[\hbox{const.}/(n_F V)]$, i. e. it
decreases {\it exponentially} with small $V$.
On the other hand, the AF gap equation, \eqref{eqaf},  {\it does} depend
on $U$. In fact, for large $U$, $E(\vk) \approx \Delta_U \propto U$,
which inserted in \eqref{eqaf} gives $\Delta_{mod}  \propto V$, i. e. 
$\Delta_{mod}$ behaves {\it linearly} with $V$, totally different from 
$\Delta_{SC}$.

Our ``projected'' \sof symmetry principle, thus, can provide an explanation
of the interrelation between the SC and the AF gap both qualitatively,
via the same dispersion, as well as semiquantitatively, i. e. via the
correct order of magnitude difference between the two modulations.

%%%
\section{Slave--boson approach}
%%%
\label{slav}

The Hartree--Fock treatment 
described in the previous section 
was physically motivated as a natural first step to show the connection of 
the SDW with the BCS solutions via a \sof theory.
 Our motivation for introducing the Hubbard repulsion $U$  was to
project out states containing doubly occupied sites and therefore to
account for the  
Mott gap
at half filling. However, 
within the Hartree-Fock treatment discussed above, the
Hubbard interaction 
is treated only perturbatively, which might be questionable for large
values of $U$.
For this reason, we repeat the calculation
using an extension of the Kotliar-Ruckenstein
 slave--boson approach \cite{ko.ru.86,be.prl,cc.prb,ex.prb}
which, while still requiring a mean-field
approximation, is more appropriate to deal with the strong repulsion
$U$.

Confidence in this approximation derives from various observations: 
(i) This method is known to give
  rather satisfying
  agreement with QMC results over a wide
  range of Coulomb correlations $U$ and values for the doping \cite{li.mu.90}. 
(ii)
  The two main results of our study, namely, that the AF gap modulation is
  correlated to the SC gap by symmetry and that the Hubbard term $H_{U}$
  induces the order of magnitude difference in these effects should be
  independent of specific approximations, at least {\it qualitatively}.
  This will be explicitly verified by comparing with the 
 simple Hartree--Fock study described above.

%%%
%%%
%%%

Within the
  Slave--boson formalism for the Hubbard model 
one introduces four bosons 
$e_\vr^{(\dagger)}$, 
$p_{\vr,\sigma}^{(\dagger)}$, and
$d_\vr^{(\dagger)}$, 
for each lattice site $\vr$, 
corresponding to empty, 
singly occupied site with spin $\sigma$, and to doubly-occupied sites, 
respectively.
Creation and annihilation of a fermion 
$c_{\vr,\sigma}$
is mapped onto the creation and annihilation of a
pseudo fermion 
$f_{\vr,\sigma}$
with an appropriate 
boson, according to the mapping:
\bea
c_{\vr,\sigma}^\dagger & \longrightarrow & 
f_{\vr,\sigma}^\dagger z_{\vr,\sigma}^\dagger
\nonumber \\
c_{\vr,\sigma}^{} & \longrightarrow & 
z_{\vr,\sigma}^{} f_{\vr,\sigma}^{} \;.
\label{equmixrep}
\eea
Kotliar and Ruckenstein have shown that, while a certain degree of
arbitrariness is allowed in the choice of 
the bosonic operator $z_\vr$,
the choice
\bea
\label{zkr}
z_{\vr,\sigma} & =& \left( 1 - d_\vr^\dagger d_\vr^{} - p_{\vr,\sigma}^\dagger p_{\vr,\sigma} 
\right)^{-\frac{1}{2}}  % \n & &
\left( e_\vr^\dagger p_{\vr,\sigma} + p_{\vr,-\sigma}^\dagger d_\vr \right) \n & &
\left( 1 - e_\vr^\dagger e_\vr^{} - p_{\vr,-\sigma}^\dagger p_{\vr,-\sigma} \right)^{-\frac{1}{2}}
\, ,
\eea
turns out to give the correct 
 free fermion limit for $U \rightarrow 0$ 
at the mean-field level.
In addition, it was shown~\cite{ex.prb} that 
\eqref{zkr} gives the correct non-interacting limit
at all 
orders in the fluctuation ($1/N$) expansion as well.
Restriction to the {\it physical} Hilbert space is achieved by
 projecting out unphysical states. 
In a functional--integral formalism this is taken care of 
 by integrating over {\it Lagrange} multipliers
$\lambda_\vr^{(1)}, \lambda_{\vr,\sigma}^{(2)}$ for each site $\vr$
\cite{ko.ru.86}. 
The total effective Hamiltonian is thus given by 
\eqref{htot}, whereby the Fermi fields have undergone the
transformation \eqref{equmixrep}, plus the Lagrange multiplier terms: 
\beqn
&&
H_{eff} = H_{tot} + \sum_{\vr} \ i \ \lambda^{(1)}_{\vr} \ 
\left( e_\vr^\dagger e_\vr^{} +  \sum_{\sigma} p_{\vr,\sigma}^\dagger
  p_{\vr,\sigma} +
d_\vr^\dagger d_\vr^{} -1
\right) 
\nonumber \\ &&
+
\sum_{\vr,\sigma} \ i \ \lambda^{(2)}_{\vr,\sigma} 
\left( f_{\vr,\sigma}^\dagger f_{\vr,\sigma} - 
     p_{\vr,\sigma}^\dagger  p_{\vr,\sigma} -
d_\vr^\dagger d_\vr^{}
\right)
\eeqn

\subsection{Antiferromagnetic phase}
\label{sbaf}

At the lowest order, the functional integral can be treated by
means of a 
 saddle--point approximation, whereby the bosonic fields are replaced by 
 time-independent complex numbers.
%%%
%%%
%%%
In the antiferromagnetic phase,
one has the usual  two-sublattices saddle point solution, i.e.
\bea 
\label{afmf}
d_{\vec r}^{}(\tau) & = & d \, , \n
e_{\vec r}^{}(\tau) &=& e \, , \n
p_{{\vec r},\sigma}^{}(\tau) &=& p_1 + e^{i \vec Q {\vec r}} \sigma p_2 \, , \n
\lambda_{\vec r}^{(1)} &=& i \lambda_1 \, , \n
\lambda_{{\vec r},\sigma}^{(2)} &=& i \lambda_2 + i e^{i \vec Q {\vec r}} \sigma \lambda_3 \, ,
\eea
from which it follows that
\be
z_{{\vec r},\sigma}(\tau) = z_1 + e^{i \vec Q {\vec r}} \sigma z_2 \;,
\ee
where
$d, e, p_1, p_2, \lambda_1, \lambda_2, \lambda_3, z_1, z_2$
can be chosen to be real numbers.

The effective slave-boson mean-field Hamiltonian  becomes
\bea
H_{\SBMF} &=& N \lambda_1 e^2 - N \lambda_1 + 2 N (p_1^2 + p_2^2)
(\lambda_1 - \lambda_2) \n & &
- 4 N \lambda_3 p_1 p_2 
+ \lambda_1 N d^2 + N U d^2 - 2 N \lambda_2 d^2
\n & &
+  \sum_{\vk,\sigma} f^\dagger_{\vk,\sigma} ( \lambda_2 - \mu ) f_{\vk,\sigma}^{}
+ \lambda_3 \sum_{\vk,\sigma} \sigma f_{\vk+\vec Q,\sigma}^\dagger f_{\vk,\sigma}^{}
\n & &
+(z_1^2 + z_2^2) \sum_{\vk,\sigma} \eps(\vk) f_{\vk,\sigma}^\dagger
f_{\vk,\sigma}^{} + H_{\sof,I} \;.
\label{hsbmf}
\eea

The slave-boson approach can only be applied to deal with the on-site repulsion 
part $U$, while the \sof part of the interaction
[$H_{\sof,I}$, cf. \eqref{equsofint}] will still be treated by the
Hartree-Fock decoupling, as in \eqref{equmfhamil}.
This is appropriate, since the \sof interaction is assumed to be
relatively weak (of the order of $J$) in contrast to
 the large Hubbard repulsion $U$. 
In our procedure, we 
{\it first} carry out the Hartree-Fock decoupling 
of the \sof interaction part of the Hamiltonian, and {\it then} apply the
slave-boson mapping \eqref{equmixrep}.
In this way, in the AF phase, we have
\beq
\label{hsofsb}
H_{\sof,I}
= \sum_{\vk,\sigma} \Delta_{\vk}^{(4)} \sigma
  c^{\dagger}_{\vk, \sigma} c^{}_{\vk+\vec Q, \sigma} \;,
\eeq
with 
\be
  \Delta_{\vk}^{(4)} 
= - \frac{V}{2\ N} |d_{\vk}| 
  \sum_{\vkp,\sigma} |d_{\vkp}| \sigma 
  \langle c^\dagger_{\vkp,\sigma} c^{}_{\vkp+\vec Q,\sigma} \rangle \, .
\ee
The transformation, \eqref{equmixrep}, with the saddle-point values
\eqref{afmf} can then be applied to \eqref{hsofsb} by first
transforming into real space. 
Going back to momentum space
 yields
\bea
H_{\sof,I} &=&
\sum_{\vk,\sigma} \Delta_\vk^{(4)} \sigma 
\left\{ (z_1^2 + z_2^2) f_{\vk,\sigma}^\dagger f_{\vk+\vec Q,\sigma}^{} \right. \n & &
\left. +
2 z_1 z_2 \sigma f_{\vk,\sigma}^\dagger f_{\vk,\sigma}^{} \right\}
\label{so5hamil}
\eea

We can now insert the \sof interaction in \eqref{so5hamil}
into the Slave--boson Hamiltonian \eqref{hsbmf},
and obtain
\bea
H_{\SBMF} &=& 
 \sum_{\vk,\sigma} f^\dagger_{\vk,\sigma} f^{}_{\vk,\sigma}
 \nonumber \\ && \times
 \left[
\eps(\vk) (z_1^2 - z_2^2) - \mu + \lambda_2 + 2 \Delta_\vk^{(4)} z_1 z_2 \right]
\nonumber \\
& + & \sum_{\vk,\sigma} f^\dagger_{\vk,\sigma} f^{}_{\vk+\vec Q,\sigma} 
\left[ \lambda_3 \sigma + \Delta_\vk^{(4)} \sigma (z_2^2 + z_1^2) \right]
+H_{bos} \nonumber \\
&=& \sum_{\vk,\sigma}^{\AF} \tvec{f_{\vk,\sigma}}{f_{\vk+\vec Q,\sigma}}^\dagger
    \sqmat{\eta_1(\vk)}{\eta_2(\vk)}{\eta_2(\vk)}{\eta_3(\vk)}
    \tvec{f_{\vk,\sigma}}{f_{\vk+\vec Q,\sigma}} 
\nonumber \\ &&
+ H_{bos} \;,
\label{mathamilaf}
\eea
with
\bea
\eta_1(\vk) &=& \eps(\vk) (z_1^2 - z_2^2) - \mu + \lambda_2 + 
                  2 \Delta_\vk^{(4)} z_1 z_2 
\n
\eta_2(\vk) &=& \lambda_3 \sigma + \Delta_\vk^{(4)} \sigma (z_2^2 + z_1^2)
\n
\eta_3(\vk) &=& -\eps(\vk) (z_1^2 - z_2^2) - \mu + \lambda_2 + 
                  2 \Delta_\vk^{(4)} z_1 z_2 \;,
\eea
where $H_{bos}$ is a purely bosonic part.
Diagonalization of \eqref{mathamilaf} yields the eigenvalues
\bea
E_{\pm}(k) &=& \lambda_2 - \mu + 2 \Delta_\vk^{(4)} z_1 z_2 
\pm  \left\{\lambda_3^2 + \eps(\vk)^2 (z_1^2 - z_2^2)^2 
\right. \n & &
        \left. + 2 \Delta_\vk^{(4)} \lambda_3 (z_1^2 + z_2^2) 
        + (\Delta_\vk^{(4)})^2 (z_1^2 + z_2^2)^2\right\}^{\frac{1}{2}}
\n
\label{equsbafeival}
\eea
Although \eqref{equsbafeival} seems to break the particle-hole
symmetry at half filling,
 this is not the case. Indeed the self-consistent solution
yields  at half filling
$\lambda_2 - \mu= z_2=0$, which makes the expression particle-hole
symmetric.
\eqref{equsbafeival} 
describes the gap structure: Since along the magnetic zone boundary the 
tight--binding dispersion $\eps(\vk)$
is identical to zero, the dispersion along the magnetic zone boundary
reads 
 at half filling: 
\be
E_{2/1}(\vk) = \pm   \lambda_3 + \Delta_\vk^{(4)} z_1^2 \;.
\ee
Therefore, we obtain a constant part of the 
gap $\Delta_U = 2 \lambda_3$ plus a modulation
\be
\label{deltasb1}
\Delta_\vk^{(4)} z_1^2 = \Delta_{\mod} |d_{\vk}| \, ,
\ee
with 
\beq
\label{deltasb2}
  \Delta_{\mod} 
  = - \frac{ V}{2 \ N \, z_1^2} \sum_{\vk,\sigma} |d_{\vk}|
  \sigma \langle c^\dagger_{\vk, \sigma} c^{}_{\vk+\vec Q,\sigma} \rangle \, .
\eeq
%%%
%%%

We introduce the unitary matrix $U(\vk)$ that diagonalizes
the Hamiltonian \eqref{mathamilaf}:
\beq
U(\vk) \sqmat{\eta_1(\vk)}{\eta_2(\vk)}{\eta_2(\vk)}{\eta_3(\vk)}
U(\vk)^{\dag} = 
\sqmat{E_{-}(\vk)}{0}{0}{E_{+}(\vk)} \;.
\eeq
The self-consistent gap equation  is given by
Eqs.~(\ref{deltasb1},\ref{deltasb2})  and takes the form (at half filling)
\bea
\Delta_{\vk}^{(4)} 
&=& -\frac{V}{2\ N} |d_{\vk}| \sum_{\vkp,\sigma} 
     |d_{\vkp}| \sigma \langle 
  c^\dagger_{\vkp,\sigma} c^{}_{\vkp+\vec Q,\sigma} \rangle \n
&=& 
 -\frac{V}{2\ N} |d_{\vk}| 
        \sum_{\vkp,\sigma}^{\AF} |d_\vkp| \sigma
\nonumber \\ && \times
\left[ U(\vk) \sqmat{2 z_1 z_2 \sigma}{z_1^2 + z_2^2}{z_1^2 + z_2^2}{2
    z_1 z_2 \sigma} U(\vk)^\dagger \right]_{1,1} \;.
\label{gapaf}
\eea

The ground--state energy 
\beq
E = H_{bos} + \sum_{\vk,\sigma}^{\AF} E_1(\vk) 
\eeq
then has to be minimized with respect to the saddle-point values of
the fields
$d, e, p_1, p_2, \lambda_1, \lambda_2$  and $\lambda_3$.
 Along with the gap--equation
Eq.~(\ref{gapaf}) this gives 8 coupled equations that are readily 
solved numerically.

%%%
\subsection{SC phase}
%%%
\label{sbsc}

The Slave--boson calculation in the superconducting
 phase is very similar to (in fact simpler than)
 the one in the AF phase. Therefore, we will skip details and only show
the main differences.
Indeed, in this case  we can use the paramagnetic {\it ansatz}
for the bosons
which avoids the complication of two different sublattices:
\bea
  d_r^{}(\tau) & = & d \n
  e_r^{}(\tau) &=& e \n
  p_{r,\sigma}^{}(\tau) &=& p \n
  \lambda_r^{(1)} &=& i \lambda_1 \n
  \lambda_{r,\sigma}^{(2)} &=& i \lambda_2  \, .
\eea
 $z_{r}$ is obviously independent of $\sigma$ and is given again by
\eqref{zkr}.
The saddle--point approximation yields the mean-field Hamiltonian
(the counterpart of \equ{hsbmf}):
\bea
  H_{\SBMF} &=& N \lambda_1 e^2 - N \lambda_1 + 2 N p^2
  (\lambda_1 - \lambda_2)  \nonumber \\
  &+& N d^2 ( \lambda_1 - 2 \lambda_2 + U )
\nonumber \\ &&
  +  \sum_{{\vec k},\sigma} f^\dagger_{{\vec k},\sigma}
    ( \eps({\vec k}) z^2 + \lambda_2 - \mu ) f_{{\vec k},\sigma}^{}
+H_{\sof,I}
 \, .
  \label{hsbmfsc}
\eea
In the SC phase, 
the \sof part reads
\be
 H_{\sof,I} = 
         z^2 \sum_{{\vec k}} \Delta_{\vec k}^{(1)} 
         \left( f^{\dagger}_{{\vec k},\up} f^{\dagger}_{-{\vec k}, \down}
         - f^{}_{{\vec k},\up} f^{}_{-{\vec k}, \down} \right) \, ,
  \label{so5hamilsc}
\ee
with the  condition
\be
  \Delta_{\vec k}^{(1)} 
  = -\frac{V}{2 N} 
d_{\vec k}  \sum_{\vkp} d_{\vkp}
  \langle c_{{\vkp},\up}^\dagger c_{-{\vkp},\down}^{\dagger} - 
   c_{{\vkp},\up}^{} c_{-{\vkp},\down}^{} \rangle \;.
\ee
The purely fermionic part 
\bea
  H_F &=& \sum_{{\vec k},\sigma} f^\dagger_{{\vec k},\sigma}
    [ \eps({\vec k}) z^2 + \lambda_2 - \mu ] f_{{\vec k},\sigma}^{} \n
    &+& z^2 \sum_{{\vec k}} \Delta_{\vec k}^{(1)} 
        \left( f^{\dagger}_{{\vec k},\up} f^{\dagger}_{-{\vec k}, \down}
         - f^{}_{{\vec k},\up} f^{}_{-{\vec k}, \down} \right) \, 
\eea
can be diagonalized by the usual Bogoliubov transformation.
This  yields the  eigenvalues
\be
  \pm E({\vec k}) = \pm \sqrt{(\eps({\vec k}) z^2 + \lambda_2 - \mu )^2+
       (\Delta_{\vec k}^{(1)})^2} \, .
\ee

Calculating the expectation value of the 
order parameter in the ground state, constructed from the
 Bogoliubov eigenstates, gives a 
self-consistent gap equation in the usual form:
\be
  1 = \frac{V}{2\ N} z^4 \sum_{\vec k} d_{\vec k}^2 \frac{1}
          {\sqrt{( \eps({\vec k}) z^2 + \lambda_2 - \mu )^2+\Delta_{\SC}^2 d_{\vec k}^2 z^4}} \;,
\ee
where $\Delta_{\SC}$ is the amplitude of the modulation
\be
  \Delta_{\vec k}^{(1)} = d_{\vec k}
 \Delta_{\SC} \, .
\ee

\begin{figure}
\begin{center}
\epsfig{file=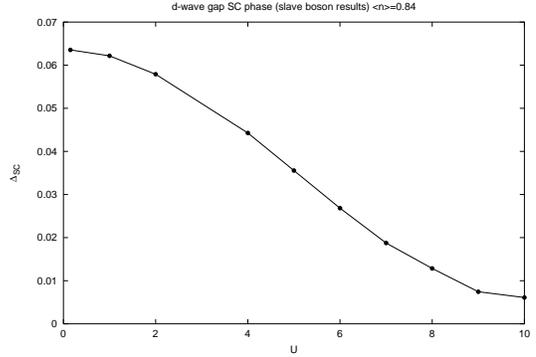, height=7cm, angle=270}
%%%
\caption{The $d$--wave gap $\Delta_{\SC}$ in the superconducting phase 
with density $\langle n \rangle = 0.84$ as a
function of the Hubbard interaction $U$
for fixed $V=0.61 \ t$.
 We see that, in contrast to the antiferromagnetic
phase, $U$ lowers the value of the gap.}
\label{fgapsc}
\end{center}
\end{figure}

The gaps obtained from 
 the slave-boson calculation are reported in
Figs.~\ref{fgapafu}, and \ref{fgapsc}. 
%%%
%%%
Figure~\ref{fgapafu} 
%%%
%%%
%%%
%%%
%%%
shows the behavior of the AF gap modulation as a function of $U$.
Just like in the simple Hartree--Fock type of mean field described in
Sec.~\ref{hart},
we observe that increasing the Hubbard--interaction
$U$  enhances the $d$--wave--like modulation $\Delta_{\mod}$
of the AF gap at half filling, and, eventually, saturates around 
$U/t\approx 8-10$. The {\it dispersion} of the gap is of course
identical to the Hartree--Fock mean-field result, 
Fig.~\ref{figafgap}.

The superconducting phase
has again a 
$d$--wave structure. 
However, in contrast to the simple Hartree-Fock mean field,
where $U$ was of no influence,
the  Hubbard interaction $U$ 
results in a {\it suppression} of
the superconducting gap $\Delta_{\SC}$ at finite doping,
as shown in
Fig.~\ref{fgapsc}.
This is due to the well-known reduction of the effective hopping
produced by the slave-boson formulation.
The combination of these two effects with increasing $U$
 (enhancement of $\Delta_{\mod}$ in the AF phase and
suppression of $\Delta_{\SC}$ in the SC phase)
yields a
ratio of the two gaps $\Delta_{\mod}/\Delta_{\SC}$ of the order of $10$,
slightly larger  than the
Hartree-Fock calculation, and more
 in agreement with  experimental findings. 

%%%

\section{Summary}
\label{summ}

%%%
%%%
%%%
%%%
%%%
%%%
%%%
%%%
%%%
%%%
%%%
%%%
%%%
%%%
%%%
%%%
%%%
%%%
%%%
%%%
%%%

Summarizing,  
we have shown that the $|d|$-wave-like modulation observed in the AF
gap at half filling in  \cacuocl, similar to the 
dispersion of the 
SC gap at finite doping, 
 indicates an intimate 
relationship between the two phases, as suggested by the \sof theory
of high-T$_c$ superconductivity.
The idea is that one gap can be mapped into the other by a \sof
transformation. This is done by using an effective \sof-invariant 
Hamiltonian which allows for such a mapping. 
%%%
%%%
However, we have also
shown that it is important to
break this symmetry by introducing
 ``by hand'' an  Hubbard interaction term in order 
 to correctly obtain 
the constant part
of the AF gap.
Via this ``projected'' \sof theory, we can interpret
the experimentally observed
 $d$--wave--like modulation of the AF gap as the fingerprint of
the superconductor in the AF state just like
the neutron resonance mode~\cite{ro.re.91,mo.ye.93,fo.ke.95} 
may be viewed as the fingerprint of the AF
correlation in the SC state~\cite{de.zh.98}.

The projection can also explain
the experimentally observed order of magnitude difference between
 the $d$--wave--like gap modulation in the AF state 
($\approx J \sim 0.12 eV$) 
and the $d$--wave gap in the SC state ($\approx J/10$).
 This different magnitude of the two gap modulations also
suggests a reason why the \neel
temperature differs so much from the superconducting transition temperature.

Finally, we have shown that the projected \sof theory can provide
a qualitative explanation
for the observation~\cite{pa.da.01}, 
 that the maximum \tc observed in a large variety
of high-\tc cuprates scales with the next-nearest-neighbor hopping
matrix element $t'$.

%%%
%%%
%%%
%%%
%%%
%%%
%%%
%%%
%%%
%%%
%%%
%%%
%%%
%%%
%%%
%%%
%%%
%%%
%%%
%%%
%%%
%%%

\section*{Acknowledgments}

We thank Z. C. Zhang, R.B.\ Laughlin, D.J.\ Scalapino, R.\ Eder, Z.-X.\ Shen,
J.\ C.\ Campuzano and O.K.\ Andersen for helpful
discussions and suggestions.
%%%
%%%
%%%
%%%
%%%
We acknowledge financial
support
by the DFG [HA 1537/17-1, HA1537/16-2, and 
partially from the Heisenberg program 
AR 324/3-1 (EA)], KONWIHR OOPCV,
and BMBF (05SB8WWA1). 
This work is dedicated to Professor E. M\"uller-Hartmann on the occasion of
his 60$^{th}$ birthday.

%%\bibliography{preprints,articles,mypublications}  %%c%%  %%c%% 
%%\bibliographystyle{myprsty} %%c%%  %%c%%
\ifx\undefined\andword \def\andword{and} \fi %%c%%
\ifx\undefined\submittedto \def\submittedto{submitted to } \fi %%c%%
\ifx\undefined\toapperarin \def\toappearin{to appear in } \fi %%c%%
\def\nonformale#1{#1}
\def\formale#1{}
\def\spa{} \def\spb{}
\spa

\appendix

\section{Details of the mean-field calculations}
\label{appa}

%%%
%%%
\subsection{BCS solution of the SC state}
%%%
\label{bcs}

At finite doping, the symmetry between SC and AF is broken to favor SC.
Therefore, we look for a BCS solution given
by a finite mean-field expectation value of the operator $\Delta$ in
\eqref{Delta}. The Hamiltonian, \eqref{htot}, becomes quadratic in the
fermionic operators and can be solved by the usual Bogoliubov
transformation. 
The mean-field condition  results in the BCS gap equation 
for the amplitude of the order parameter $\Delta_{SC}$, i. e.
\be
\label{equbcsgap}
  1 = \frac{V}{N} \sum_{\vk} \frac{d_{\vec k}^2}{2 \sqrt{(\eps_{\vk} - \mu)^2 +
\Delta_{\SC}^2 d_{\vec k}^2}} \;,
\ee
which has to be solved in a self--consistent way for a given value of
the parameter $V$. 
Notice that the on-site interaction
 $U$ does not enter this equation, due to the 
fact that, in contrast to $s$-wave,
 two holes of a $d$-wave pair never occupy the same site.

The $k$-dependent  SC gap 
is given by $\Delta_{SC} d_{\vec k}$ and depends on two parameters:
the magnitude $\Delta_{SC}$ and the parameter $b$ introduced in
\eqref{dgen}, giving its shape.
Our procedure consists in fitting these two parameters to the recent data
on the SC gap by Mesot et al.~\cite{me.no.99}.
In that paper, the SC gap obtained by ARPES experiments was fitted
with a $d$-wave form including nearest- and
next-nearest-neighbor terms:
\beq
\label{dnnn}
d_{\vec k} =  b (\cos k_x - \cos k_y) + (1-b) 
(\cos 2 k_x - \cos 2 k_y) \;.
\eeq
This form  cannot be 
made consistent with our \sof hypothesis, since it would require a
form factor \eqref{ggen} without the property $g_{\vec k + \vec Q} = - 
g_{- \vec k}$.
However, a fit of the SC gap by the form, \eqref{dgen}, turns out to be
as good as one from the next-nearest-neighbor form, \eqref{dnnn}.

Our best fit to the data of Ref.~\onlinecite{me.no.99} with \eqref{dgen} gives
$b= 0.81$, and $\Delta_{SC} = 0.04 t$.  
The values of the parameters can be inserted in \eqref{equbcsgap}
to obtain the appropriate value of $V$ ($V=0.89 t$), %%??thomas
 which can now be used to study 
the AF phase.
%%%
%%%

%%%
%%%
\subsection{Hartree-Fock solution for the Antiferromagnetic phase}
%%%
\label{hart}

We now look for the AF solution of the mean-field equation for the
Hamiltonian, \eqref{htot}, at half filling.
This is slightly more complicated than the BCS result,
\eqref{equbcsgap}, due to the interplay between the \sof interaction
in \eqref{hso5} and the Hubbard repulsion term.

We introduce a momentum dependent 
SDW operator polarized in the $z$--direction:
\be
 n_{\vk}  =  
c^\dagger_{\vk + \vec Q} \sigma^z
c_{\vk} = 2 N_3(\vk) \, .
\ee
Within a mean-field decoupling, the Hamiltonian in \eqref{htot} 
becomes
\bea
H_{\MF} &=& \sum_{\vk,\vec \sigma} \eps(\vk) c^\dagger_{\vk,\sigma} 
c^{}_{\vk, \sigma} 
- \frac{U}{2 \, N} [ \sum_{\vkp} \langle n_{\vkp} \rangle ]
\sum_{\vk} n_{\vk} 
 \n & &
 -\frac{V}{2\ N}  [\sum_{\vkp} |d_{\vkp}| \langle n_{\vkp} \rangle ]
     \sum_{\vk} |d_{\vk}| n_{\vk} 
  \n & &
 = \sum_{\vk,\vec \sigma} \eps(\vk) c^\dagger_{\vk,\sigma} c^{}_{\vk, \sigma}
    - \Delta_U  \sum_{\vk} n_{\vk}
\n & &
- \Delta_{\mod} \sum_{\vk} |d_{\vk}| n_{\vk} \;,
\label{equmfhamil}
\eea
with the self-consistent parameters
\bea
\label{deltaudeltamod}
  \Delta_U &=&  \frac{U}{2 \, N} [ \sum_{\vkp} \langle n_{\vkp} \rangle ]\, ,\\
  \Delta_{\mod} &=& \frac{ V}{2\, N}
               [\sum_{\vkp} |d_{\vkp}| \langle n_{\vkp} \rangle ] \, .
\eea

It is convenient to recast
\equ{equmfhamil}
 in matrix form
as a sum over the 
antiferromagnetic Brillouin zone (AFBZ), 
\be
H = \sum_{\vk,\sigma}^{AFBZ} \tvec{c_{\vk,\sigma}}{c_{\vk+\vec Q,\sigma}}^\dagger
    \cdot M(\vk) \cdot  
    \tvec{c_{\vk,\sigma}}{c_{\vk+\vec Q,\sigma}} \;,
  \label{equhfmfhamil}
\ee
with
\be
  M(\vk) = \sqmat{\eps(\vk)}{-\sigma(\Delta_U +\Delta_{\mod} |d_{\vk}|)}
          {-\sigma(\Delta_U +\Delta_{\mod} |d_{\vk}|)}{-\eps(\vk)} \, .
\ee
The Hamiltonian (\equ{equhfmfhamil}) can be diagonalized by the usual 
transformation into Bogoliubov operators
\bea
\gamma_{\vk,\sigma}^c &=& u_{\vk} c_{\vk,\sigma} + \sigma v_{\vk} 
  c_{\vk+\vec Q,\sigma} \, , \n
\gamma_{\vk,\sigma}^v &=& v_{\vk} c_{\vk,\sigma} - \sigma u_{\vk} 
  c_{\vk+\vec Q,\sigma} 
\eea
with  amplitudes
\bea
  u_{\vk} = \frac{1}{\sqrt{2}} \sqrt{ 1 + \frac{\eps(\vk)}{E(\vk)} } \, , \n
  v_{\vk} = \frac{1}{\sqrt{2}} \sqrt{ 1 - \frac{\eps(\vk)}{E(\vk)} } \, , \n
\eea
and eigenvalues
\be
\label{ek}
  \pm E(\vk) = \pm 
     \sqrt{ \eps(\vk)^2 + (\Delta_U + |d_{\vk}| \, \Delta_{\mod})^2} \, .
\ee
 At half--filling, the ground state consists of all states of $\gamma^c$ 
particles occupied ($c$ stands for
{\it conduction} band with $E(\vk) < 0$ and $v$ for {\it valence} band with
$E(\vk) > 0$) and the expectation value $\langle n_{\vk} \rangle$ can
be readily evaluated.

%%%

%%%

\end{multicols}
\end{document}